\documentclass[aps,pre,epsf,preprint]{revtex4}

\usepackage{graphicx}

\begin{document}  
\draft

\title{Effect of water-wall interaction potential on the properties of nanoconfined water}

\author{Pradeep Kumar$^1$, Francis W. Starr$^2$, Sergey V. Buldyrev$^{1,3}$, and H. Eugene Stanley$^1$}

\address{$^1$Center for Polymer Studies and Department of Physics\\
 Boston University, 590 Commonwealth Avenue, Boston, MA 02215 USA\\
$^2$Department of Physics, Wesleyan University, Middletown, CT 06459 USA\\
$^3$Department of Physics, Yeshiva University, 500 West 185th Street, New York, NY 10033 USA }

\date{last revised: \today,~~~ksbs.tex}

\begin{abstract}   

Much of the understanding of bulk liquids has progressed through study
of the limiting case in which molecules interact via purely repulsive
forces, such as a hard-core potential.  In the same spirit, we report
progress on the understanding of confined water by examining the
behavior of water-like molecules interacting with planar walls via
purely repulsive forces and compare our results with those obtained
for Lennard-Jones (LJ) interactions between the molecules and the
walls. Specifically, we perform molecular dynamics simulations of 512
water-like molecules which are confined between two smooth planar
walls that are separated by 1.1~nm. At this separation, there are
either two or three molecular layers of water, depending on density.
We study two different forms of repulsive confinements, when the
interaction potential between water-wall is (i) $1/r^9$ and (ii)
WCA-like repulsive potential. We find that the thermodynamic, dynamic
and structural properties of the liquid in purely repulsive
confinements qualitatively match those for a system with a pure LJ
attraction to the wall. In previous studies that include attractions,
freezing into monolayer or trilayer ice was seen for this wall
separation.  Using the same separation as these previous studies, we
find that the crystal state is not stable with $1/r^9$ repulsive walls
but is stable with WCA-like repulsive confinement. However, by
carefully adjusting the separation of the plates with $1/r^9$
repulsive interactions so that the effective space available to the
molecules is the same as that for LJ confinement, we find that the
same crystal phases are stable.  This result emphasizes the importance
of comparing systems only using the same effective confinement, which
may differ from the geometric separation of the confining surfaces.

\end{abstract}

\maketitle

\section{Introduction}

Confinement of water in nanopores affects many properties of water,
such as freezing temperature, crystal
structure~\cite{bellisent1,koga1,koga2,koga2005,zangi-rev,zangi1}, the
glass transition temperature, and the position of the hypothesized
liquid-liquid (LL) critical
point~\cite{zanotti1999,ChenPRVT,bellisent,xu2005pnas,ku2005sub}.
Indeed, water confined in nanoscale geometries has received much recent
attention, in part because of its importance in biology, engineering,
geophysics and atmospheric sciences.  The effects of different kinds
of confinement have been studied, both using experiments and
simulations~\cite{antognozzi,gilijamse,seemasingh,koga1,koga2,zangi1,zangi2,kumar2005,nicolas,matri,gallo2002}.

Bulk supercooled water -- water cooled below the equilibrium freezing
temperature -- shows many anomalous
properties~\cite{Debenedetti03a,Debenedetti03b,angell-review,bellisent1}.
Experiments find that at low temperatures, various response functions,
such as isothermal compressibility and specific heat, increase
sharply. There has been comparatively less research on confined water.
Using the ST2 potential to model water confined between smooth plates
\cite{meyer}, a LL phase transition has been proposed.  A
liquid-to-amorphous transition is seen in simulations of water using
the TIP4P potential \cite{tip4p} confined in carbon nanotubes
\cite{koga2}.  Recent theoretical work \cite{truskett} suggests that
hydrophobic Lennard-Jones (LJ) confinement shifts the LL transition to
lower temperature and lower pressure compared to bulk water, a feature
also found in simulations of water confined between hydrophobic
plates~\cite{kumar2005}.

Confinement is known to enhance solidification of molecules that are
more or less spherical~\cite{rhykerd,brushan,persson}. However,
careful experiments on thin films of water show that water performs
extremely well as a lubricant, suggesting that confined water may be
more fluid than bulk water~\cite{fluidity}.  Recent experiments show
that water in hydrophilic confinement, when cooled to very low $T$,
does not freeze~\cite{bellisent} -- a phenomenon also supported by
simulation studies~\cite{Gallo00,Gallo00b}.  In contrast,
simulations~\cite{koga1,koga2,zangi1,zangi2,kumar2005,nicolas} show
that hydrophobically confined water does freeze into different
crystalline structures, which do not have counterparts in bulk water.
Indeed monolayer, bilayer and trilayer ice have all been found in
simulations~\cite{koga1,koga2,zangi1,zangi2,kumar2005,nicolas}.  Thus
hydrophobic confinement seems to facilitate the freezing of
water. However the reason for this facilitation is not yet fully
understood. The hydrogen-bond interaction between water molecules is
an order of magnitude stronger than Van der Waals attraction with the
hydrophobic walls. Thus one may hypothesize that freezing in
hydrophobic confinement depends critically on the separation between
confining walls that may distort or facilitate a particular
crystalline structure, rather than on the weak details of the
water-wall interaction potential. To test this hypothesis we perform
molecular dynamics (MD) simulations of water in two different forms of
repusive confinement. Specifically, we study:

\begin{itemize}

\item The $1/r^9$ repulsive part of the LJ potential studied in
Ref.~\cite{kumar2005}.

\item The same potential used in Ref.~\cite{kumar2005} but truncated
and shifted such that there is no attractive part in the potential,
analogous to the Weeks-Chandler-Anderson (WCA)
potential~\cite{weeks}. This potential allows us to examine the role,
if any, the attractive part of the water-wall LJ potential plays in
determining the thermodynamics and structure of confined water.

\end{itemize}

We compare the case when the water-wall interactions are purely
repulsive (``repulsive confinement'') with the studied case of
pure LJ confinement~\cite{kumar2005}. We also compare the freezing in
repulsive confinements with the freezing found when the water-wall
interactions are represented by an LJ interaction (``LJ hydrophobic
confinement'')~\cite{kumar2005,koga1}.

This paper is organized as follows: In Sec.~II, we provide details of
our simulations and analysis methods. Simulation results for the
liquid state are provided in Sec.~III. In Sec.~IV we discuss the
freezing properties of our system.

\section{Simulation and Analysis Methods}

We perform MD simulations of a system composed of
water-like molecules confined between two smooth walls.  The molecules
interact via the TIP5P pair potential \cite{jorgensen1} which, like
the ST2~\cite{st2} potential, treats each water molecule as a
tetrahedral, rigid, and non-polarizable unit consisting of five point
sites~\cite{note1}. The TIP5P potential predicts many of the anomalies
of bulk water \cite{masako}. For example, TIP5P reproduces the density
anomaly at $T=277$~K and $P=1$~atm and its structural properties
compare well with experiments
\cite{masako,jorgensen1,jorgensen2,sorenson,kuOdessa}.  TIP5P is known
to crystallize at high pressures~\cite{masako} within accessible
computer simulation time scales, and shows a ``nose-shaped'' curve of
temperature versus crystallization time~\cite{masako}, a feature found
in experimental data on water solutions \cite{baez}.

\begin{figure}[htb]
\includegraphics[width=10cm]{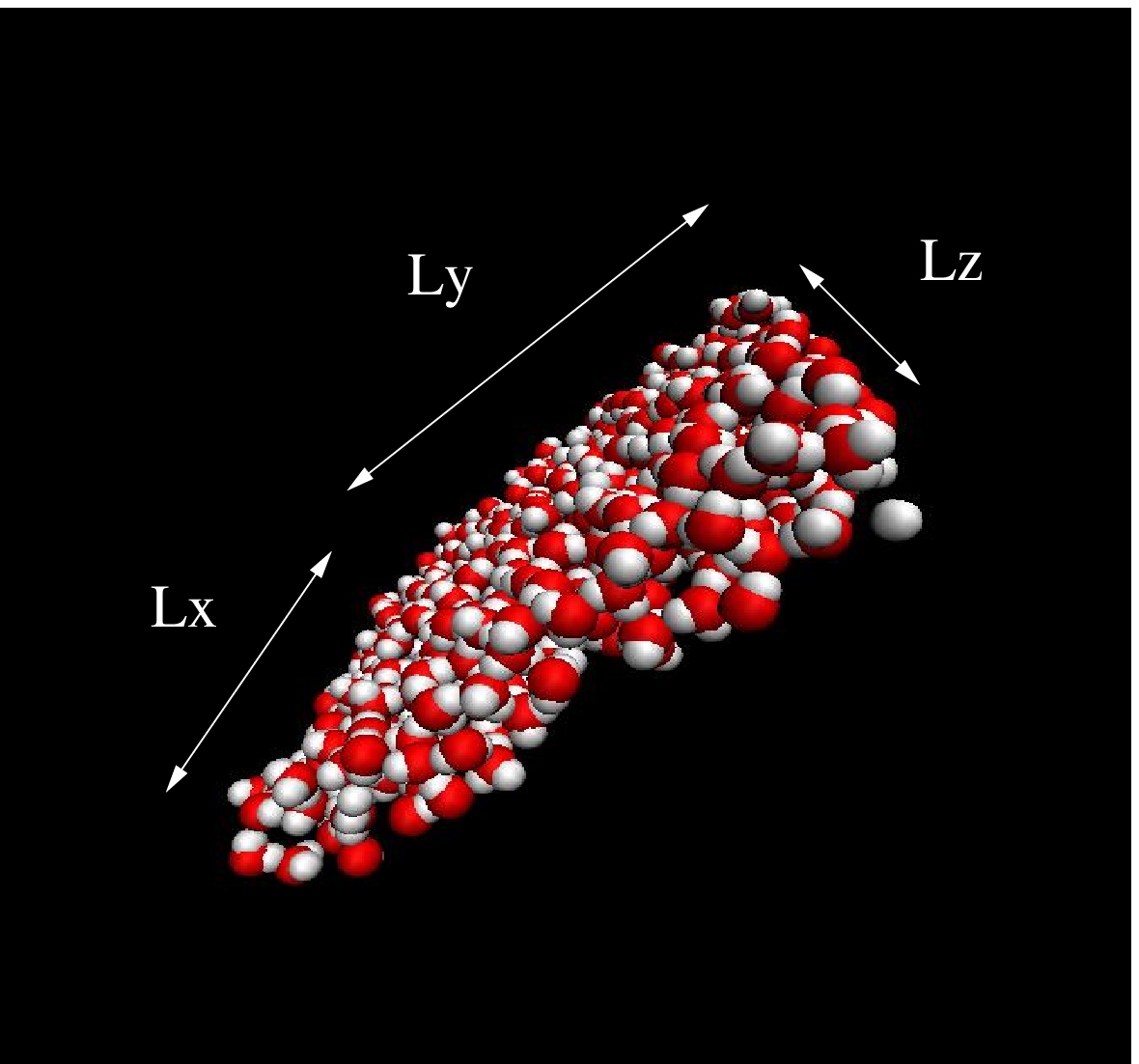}
\caption{Perspective view of the system, showing the 512 water
molecules confined between two walls perpendicular to the
z-direction. Note that the confining plates are located along the
z-direction and are separated by 2-3 molecular layers of water.}
\label{fig:schematic}
\end{figure}

In our simulations, $N=512$ water molecules are confined between two
 smooth planar walls, as shown schematically in
 Fig.~\ref{fig:schematic}.  The walls are located at $z_w=\pm0.55$~nm
 (wall-wall separation of $1.1$~nm), which results in $\approx 2-3$
 layers of water molecules.  Periodic boundary conditions are used in
 the $x$ and $y$ directions, parallel to the walls.

\begin{figure}[htb]
\includegraphics[width=10cm]{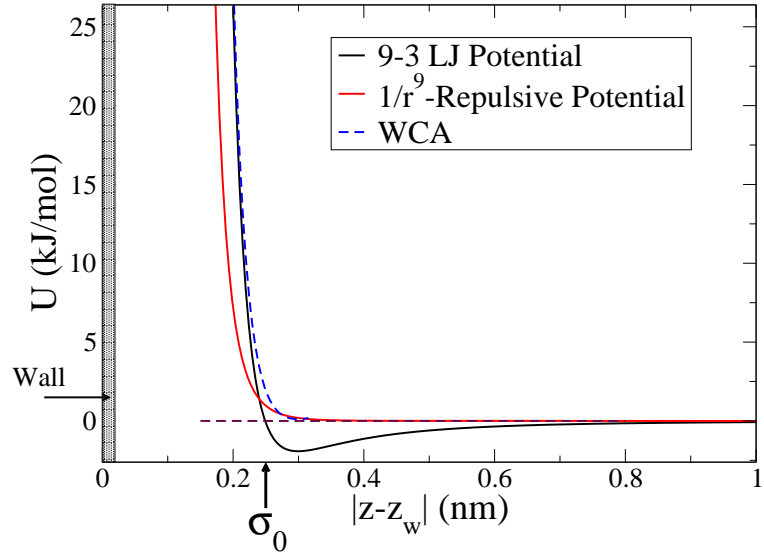}
\caption{Water-wall interaction potentials namely the 9-3 LJ,
$1/r^9$ repulsive, and WCA potential as a function of distance of
water molecules $|z-z_w|$ from the center of one of the walls (shaded
rectangle).}
\label{fig:potentials}
\end{figure}

We study two different forms of purely repulsive water-wall
interaction. The first uses only the $r^{-9}$ repulsive core, which we
call the $1/r^9$ repulsive potential,
\begin{equation}
U\left(z-z_{W}\right) =
4\epsilon_{\rm OW}\left[\left(\frac{\sigma_{\rm OW}}{|z-z_W|}\right)^9\right].
\label{Udz}
\end{equation}
Here $|z-z_W|$ is the distance from the oxygen atom of a water
molecule to the wall, while $\epsilon_{\rm OW} = 0.25$ kJ/mol and
$\sigma_{\rm OW}= 0.25$ nm are potential parameters
(Fig.~\ref{fig:potentials}). Similar but different parameter values
were used in previous confined water simulations using the TIP5P
interaction potential~\cite{kumar2005}. Specifically, in Ref.~\cite
{kumar2005}, the water-wall interaction was modelled using a 9-3
LJ potential with $\epsilon_{\rm OW}=1.25$~kJ/mol and
$\sigma_{OW}=0.25 $~nm. We choose a different $\epsilon_{\rm OW}$ in
the case of repulsive confinement so that the repulsion between the
water and wall decays to almost zero where the 9-3 LJ -potential has a
minimum.

The second purely repulsive potential uses both attractive and
repulsive terms of the 9-3 LJ potential, but truncates and shifts the
potential at the position of the minimum to create a repulsive
potential that exactly mimics the repulsion of ref.~\cite{kumar2005},
in analogy to the WCA potential,

\begin{displaymath}
U(z-z_W) = \left\{ \begin{array}{ll}
    4\epsilon_{\rm OW} [ (\frac{\sigma_{\rm OW}}{z-z_W})^9-(\frac{\sigma_{\rm OW}}{z-z_W})^3 ] + \frac{8\epsilon_{\rm OW}}{3^{3/2}} & \textrm{if $|z-z_W| < 3^{1/6}\sigma_{\rm OW}$}\\
 0 & \textrm{if $|z-z_W| > 3^{1/6}\sigma_{\rm OW}$},\\
\end{array} \right.
\end{displaymath}


%
where $\epsilon_{\rm OW}=1.25$~kJ/mol and $\sigma_{OW}=0.25 $~nm. For
the $1/r^9$ repulsive potential, we perform simulations for $56$ state
points, corresponding to seven temperatures $T=220$~K, $230$~K,
$240$~K, $250$~K, $260$~K, $280$~K, and $300$~K, and eight ``geometric
densities'' $\rho_g=0.60$~g/cm$^3$, $0.655$~g/cm$^3$,
$0.709$~g/cm$^3$, $0.764$~g/cm$^3$, $0.818$~g/cm$^3$,
$0.873$~g/cm$^3$, $0.927$~g/cm$^3$, and $0.981$~g/cm$^3$ -- the same
as studied in Ref.~\cite{kumar2005}.  The geometric values of density
do not take into account the fact that the repulsive interactions of
molecules with the walls increases the overall amount of available
space, since the $\epsilon_{OW}$ parameter of the $1/r^9$ repulsive
potential is smaller than the $\epsilon_{OW}$ used for the LJ confined
system~\ref{fig:potentials}. For systems confined by LJ interactions,
there is a well-defined preferred distance from the wall, making it
relatively straightforward to evaluate the ``effective'' density of
molecules confined by the attractive wall.  In our system with only
repulsive interactions, there is no such preferred distance, as
emphasized by Fig.~\ref{fig:potentials}.

\begin{figure}[htb]
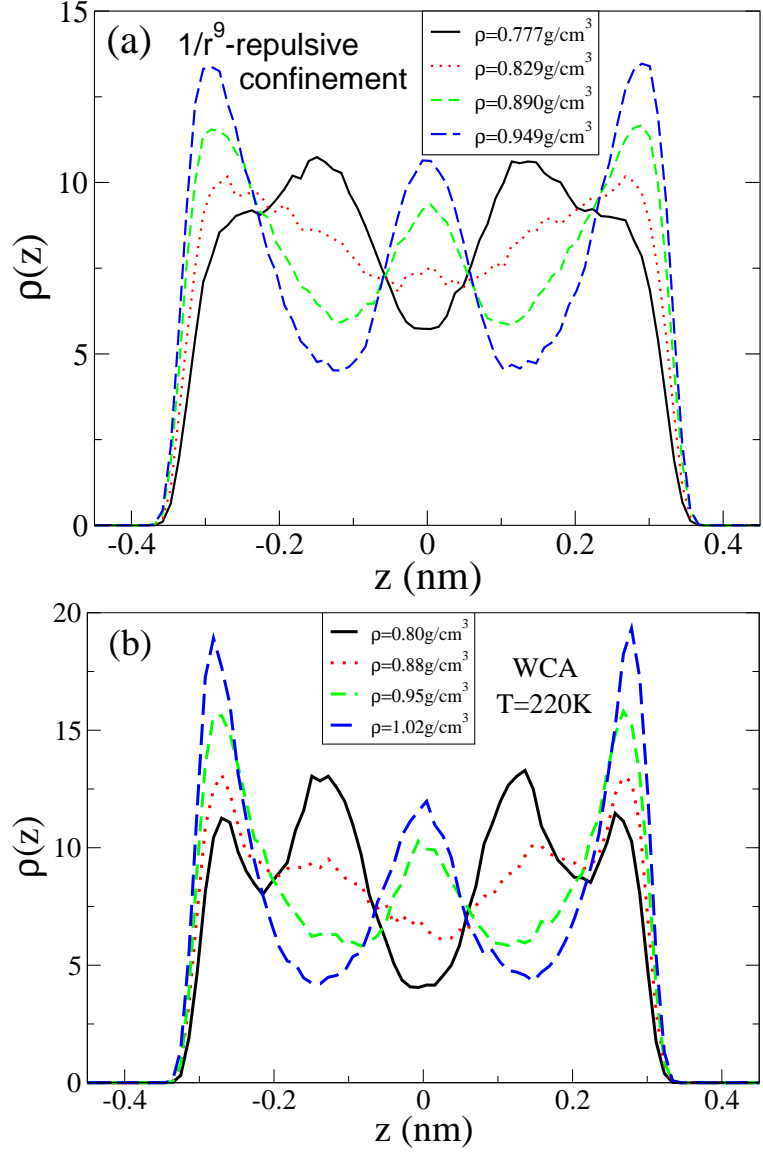

\includegraphics[width=10cm]{fig3a.eps}
\includegraphics[width=10cm]{fig3b.eps}
\caption{Density profile $\rho (z)$ along z-direction for four
different bulk densities at T=$250$~K for $1/r^9$ repulsive
confinement. (b) Density profile $\rho (z)$ along z-direction for four
different bulk densities at T=$220$~K for the case of WCA confinement. Both the repulsive confinements show similar layering of water molecules as seen in the case of LJ confinement~\cite{zangi1,kumar2005}.}
\label{fig:rhoz}
\end{figure}

We can approximate the effective density by examining the local
density $\rho(z)$ (Fig.~\ref{fig:rhoz}). We utilize the fact that
$\rho(z)$ has an inflection, and estimate the effective $L_{z}$ by the
location where the second derivative of $\rho(z) = 0$, or when first
derivative of $\rho(z)$ has a maximum.  We must also add to this value
of $L_{z}$ the molecular diameter of water(0.278~nm) to calculate the
real space available along $z$-direction.  The resulting ``effective
densities'' are $\rho=0.715$~g/cm$^3$, $0.777$~g/cm$^3$,
$0.829$~g/cm$^3$, $0.890$~g/cm$^3$, $0.949$~g/cm$^3$,
$1.000$~g/cm$^3$, $1.060$~g/cm$^3$, and $1.115$~g/cm$^3$.  We will use
these effective densities throughout the paper, since they will be
most comparable to the effective densities with LJ confinement.

For the WCA potential, we perform simulations for 32 state points,
corresponding to 8 different temperatures and 4 different ``geometric
densities'', $0.60$~g/cm$^3$,
$0.655$~g/cm$^3$,$0.709$~g/cm$^3$,$0.764$~g/cm$^3$ respectively. These
geometric densities correspond to ``effective densities''
$0.80$~g/cm$^3$, $0.88$g/cm$^3$, $0.95$g/cm$^3$, $1.02$g/cm$^3$
respectively. Note that these effective densities were calculated
using the method described in ~\cite{meyer,kumar2005}.

We control the temperature using the Berendsen thermostat with a time
constant of $5$ ps \cite{berend} and use a simulation time step of
$1$~fs, just as in the bulk system \cite{masako}. Water-water
interactions are truncated at a distance~$0.9$~nm as discussed in
Ref.~\cite{jorgensen1}.

\section{Thermodynamics and Structure}

\begin{figure}[htb]
\includegraphics[width=10cm]{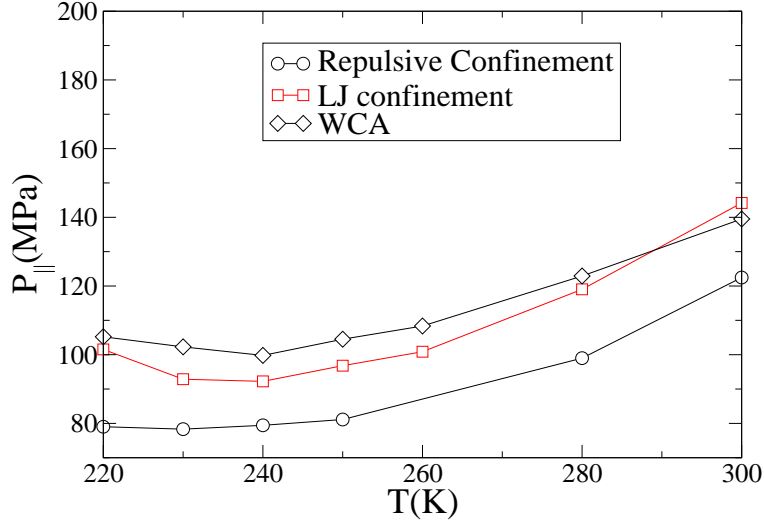}
\caption{Lateral pressure $P_{\|}$ for one isochore for the purely
 repulsive, LJ confinement and WCA cases. Here the effective density
 is $\rho=0.829$~g/cm$^3$ for the $1/r^9$ repulsive potential and
 $\rho=0.950$~g/cm$^3$ for LJ confinement. These effective densities
 for both systems correspond to the same geometric density of
 $\rho_{\rm g} = 0.709$~g/cm$^3$. All forms of confinement show a TMD, indicated by
 the minimum of the pressure; however the TMD is very ``flat'' for
 $1/r^9$ repulsive and WCA confinement. As expected the value of
 $P_{\|}$ approaches the value of $P_{\|}$ for the case of LJ confinement
 at high temperatures.}
\label{fig:pxy-comp}
\end{figure}

One of the defining characteristics of water is the existence of a
temperature of maximum density (TMD). Relative to bulk water -- LJ
confinement shifts the locus of the TMD to lower $T$ by $\approx
40$~K~\cite{kumar2005}. Additionally, the sharpness of the density
maximum is markedly decreased in comparison to the bulk.
Fig.~\ref{fig:pxy-comp} shows isochores of $P$ for LJ confinement,
$1/r^9$ repulsive confinement and WCA confinement, for similar
densities. A TMD in this plot is coincident with the minimum in the
isochore.  For $1/r^9$ repulsive confinement, the minimum is very
weak, but the location of the flatness in the isochore is near to that
of the system with LJ confinement.  This result suggests the the
$1/r^9$ repulsive confinement further suppresses the structural
ordering of the molecules that is known to be responsible to the
presence of a density maximum. The TMD for the case of WCA confinement
again appears at the same $T$ as the $1/r^9$ repulsive confinement and
LJ confinement cases but the isochore in the TMD region is flatter
than for case the of LJ confinement confinemen. Hence both kinds of
the repulsive confinement supprsess the structural ordering in lateral
directions compared to the case of bulk and LJ confinement. We further
notice that the value of the lateral pressure $P_{\|}$~in the case of
the LJ confinement approaches the value of $P_{\|}$ in case of WCA
confinement at high temperatures. This behavior of $P_{\|}$ for
LJ confinement should be expected since at very high temperatures the
molecules will not feel the potential minimum of the water-wall
interaction.

In order to compare the structural properties of repulsive
confinement with those of LJ confinement, we calculate the lateral
oxygen-oxygen radial distribution function (RDF) defined by
\begin{equation}
g_{\|}(r) \equiv \frac{1}{\rho^2 V}\sum_{i \ne j} \delta(r-r_{ij})
\left[\theta \left(|z_i-z_j|+\frac{\delta z}{2}\right) -\theta
\left(|z_i-z_j|-\frac{\delta z}{2}\right)\right].
\end{equation}
Here $V$ is the volume, $r_{ij}$ is the distance parallel to the walls
between molecules $i$ and $j$, $z_i$ is the $z$-coordinate of the oxygen
atom of molecule $i$, and $\delta(x)$ is the Dirac delta function. The
Heaviside functions, $\theta(x)$, restrict the sum to a pair of oxygen
atoms of molecules located in the same slab of thickness $\delta z =
0.1$~nm.  The physical interpretation of $g_{\|}(r)$ is that $g_{\|}(r)
2 \pi r dr \delta z$ is proportional to the probability of finding an
oxygen atom in a slab of thickness $\delta z$ at a distance $r$ (parallel
to the walls) from a randomly chosen oxygen atom. In a bulk liquid, this
would be identical to $g(r)$, the standard RDF.

\begin{figure}[htb]
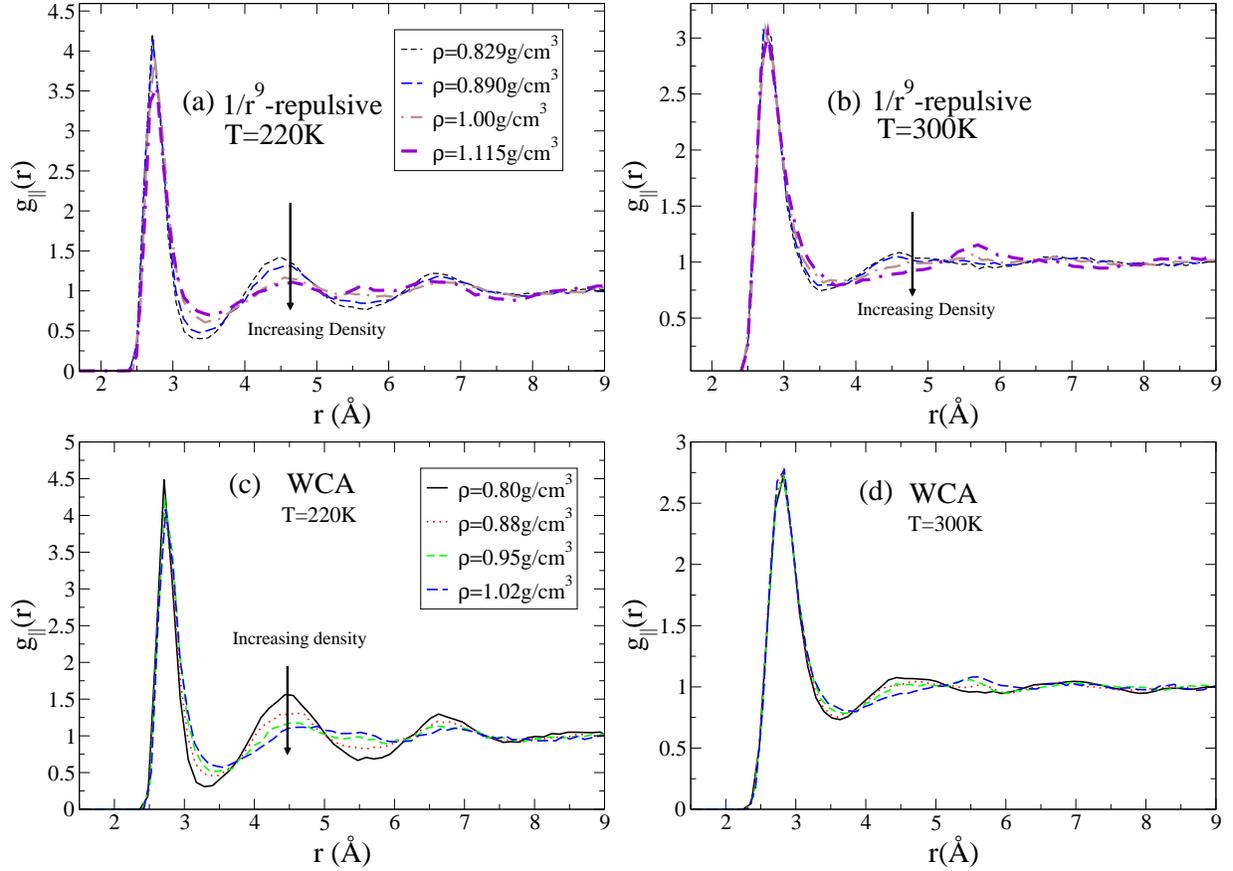

\includegraphics[width=8cm]{fig5a.eps}
\includegraphics[width=8cm]{fig5b.eps}
\includegraphics[width=8cm]{fig5c.eps}
\includegraphics[width=8cm]{fig5d.eps}
\caption{Lateral oxygen-oxygen pair correlation function $g_{\|}(r)$
   for the case of $1/r^9$ repusive and WCA confinements. Shown are
   four different densities at two fixed temperatures (a) $T=220$~K
   and (b) $T = 300$~K for $1/r^9$ repulsive confinement and (c) and
   (d) for WCA confinement. Note that with increasing density, the
   second neighbor peak at $\approx 0.45$~nm becomes less pronounced
   and at high temperatures moves to a larger distance seen in
   LJ confinement~\cite{kumar2005}. }
\label{grxy}
\end{figure}

Figure~\ref{grxy} shows the temperature and density dependence of the
lateral oxygen-oxygen pair correlation function for both $1/r^9$
repulsive (Fig.~\ref{grxy}a, Fig.~\ref{grxy}b) and WCA
(Fig.~\ref{grxy}c, Fig.~\ref{grxy}d) confinements. For both repulsive
confinements, the qualitative behavior of the dependence of $g_\|$ is
the same. At low temperature and low density, the first two peaks in
$g_{\|}$ appear at $r=2.78$~\AA and $r=4.5$~\AA, but at high densities
the second peak moves to a larger distance.  This behavior is nearly
identical to that observed for water confined between LJ surfaces, and
is discussed in detail in Ref.~\cite{kumar2005}.

\begin{figure}[htb]
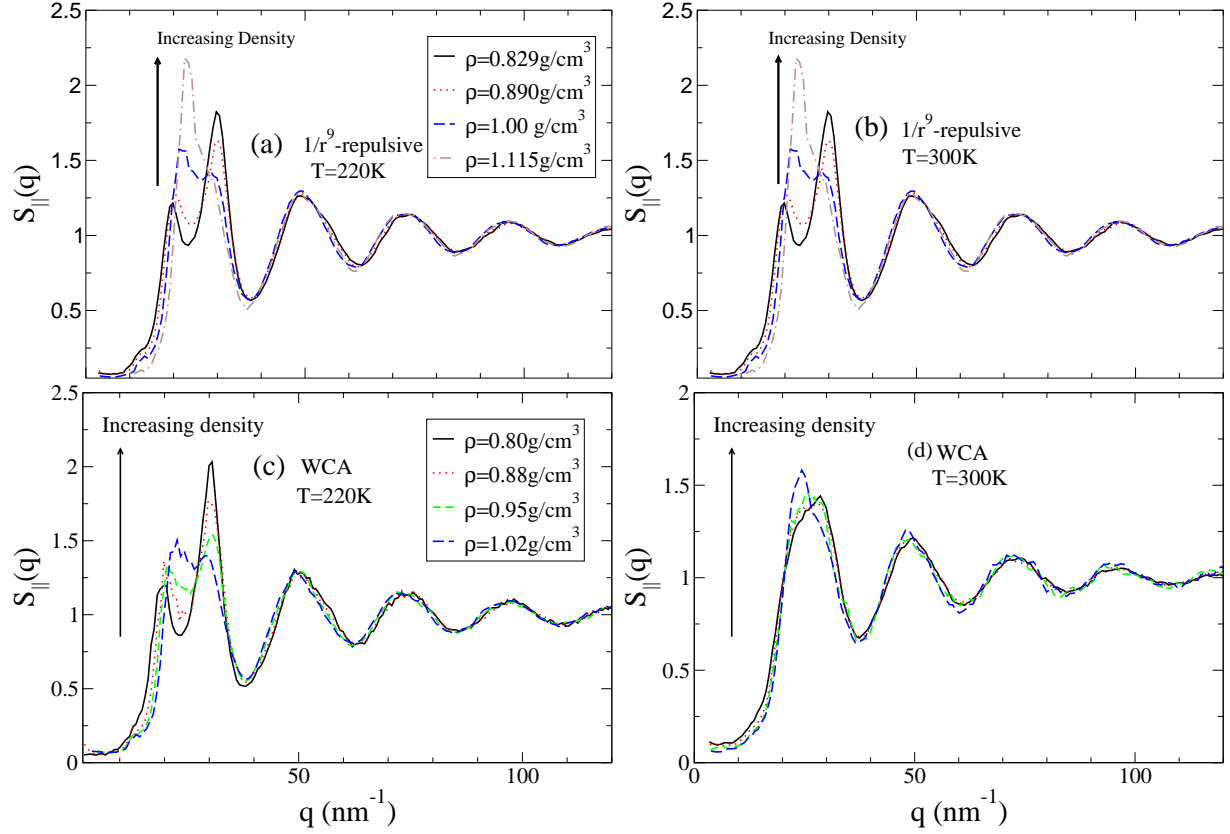

\includegraphics[width=8cm]{fig6a.eps}
\includegraphics[width=8cm]{fig6b.eps}
\includegraphics[width=8cm]{fig6c.eps}
\includegraphics[width=8cm]{fig6d.eps}
\caption{Lateral structure factor $S_{\|}(q)$ for the case of
  $1/r^9$ repulsive and WCA confinements. Shown are four different
  densities at two fixed temperatures (a) $T=220$~K and (b) $T =
  300$~K for $1/r^9$ repulsive confinement and (c) and (d) for WCA
  confinement for the same temperatures. The first peak of $S_{\|}(q)$
  corresponding to the hydrogen-bonds weakens as density is increased
  and is absent at high densities and high temperatures. The first
  peak of $S_{|}(q)$ in the case of $1/r^9$ repulsive confinement
  weaker than the LJ confinement.}
\label{fig:sqxy}
\end{figure}

\begin{figure}[htb]
\includegraphics[width=10cm]{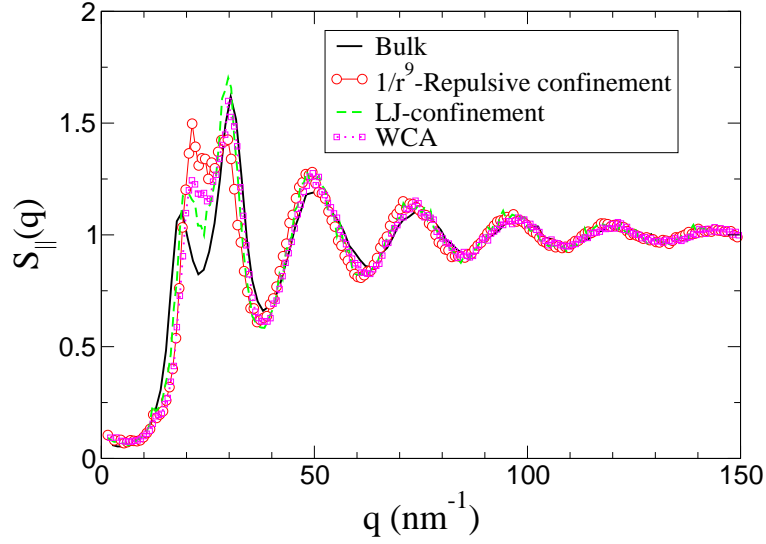}
\caption{ A comparison of the structure factors for different
confinements with bulk water structure factor at T=250K. For the
comparion we choose the effective densities in confinements close to
each other. Densities for $1/r^9$ repulsive confinement, LJ
confinement, WCA-confinement are chosen to be 0.950~g/cm$^3$. We
choose the structure factor for bulk water at density 1.00~g/cm$^3$
. A diminished peak at $\approx 18$~nm$^{-1}$ shows that the local
tetrahedral structure is weakened in case of all forms of
confinement. A further comparison of LJ confinement at 0.950~g/cm$^3$
with $1/r^9$ repulsive confinement shows that water in $1/r^9$
repulsive confinement is less tetrahedral, as the first peak of
$S_{\|}(q)$ is much weaker than the first peak of $S_{\|}(q)$ for LJ
confinement. Water in WCA confinement is more structured than $1/r^9$
repulsive confinement, but is less structured than LJ confinement.}
\label{fig:sqxy-comp}
\end{figure}

We also confirm the structural similarity with LJ confinement by
calculating the lateral static structure factor $S_{\|}(q)$, defined
as the Fourier transform of the lateral RDF
$g_{\|}(r)$,
\begin{equation}
S_{\|}\left(q\right) \equiv \frac{1}{N}\sum_{j,k}\left<e^{i\vec{\bf
q}.(\vec{\bf r_j} - {\bf \vec{r_k}})}\right>.
\end{equation}
Here the $q$-vector is the corresponding wave vector in the $xy$ plane
and $r$ is the projection of the position vector on the $xy$ plane. In
Fig.~\ref{fig:sqxy}, we show the temperature and pressure dependence
of lateral structure factors for both repulsive confinements. For both
forms of repulsive confinement, the temperature and density dependence
of $S_{\|}$ is similar. We find that confined water has a weaker
pre-peak at $\approx 18$nm$^{-1}$ compared to bulk water
(Fig.~\ref{fig:sqxy-comp}), consistent with the possibility that the
local tetrahedrality is weakened by repulsive confinements. Of the
three forms of confinement, the $S_{\|}$ for LJ confinement is most
like bulk water (Fig.~\ref{fig:sqxy-comp}). Local tetrahedrality
becomes weaker in case of repulisive confinements compared to LJ
confinement. Further, we see that the water in $1/r^9$ repulsive
confinement is less structured in lateral directions compared to water
in WCA confinement, indicated by a less sharp and broad pre-peak at
$\approx 18$~nm$^{-1}$~ in $S_{\|}(q)$~(Fig.~\ref{fig:sqxy-comp}).

\section{Freezing of TIP5P water}
\label{sec:freezing}

Bulk TIP5P water crystallizes within the simulation time for $\rho
\gtrsim 1.15$~g/cm$^3$ at low temperatures~\cite{masako}. Crystallization
of confined water is seen in some simulations
\cite{zangi1,koga1,kumar2005}.  A similar crystallization appears in
simulations when an electric field is applied in lateral directions to
a system of water confined between silica walls~\cite{zangi2}.

\begin{figure}[htb]
\includegraphics[width=12cm]{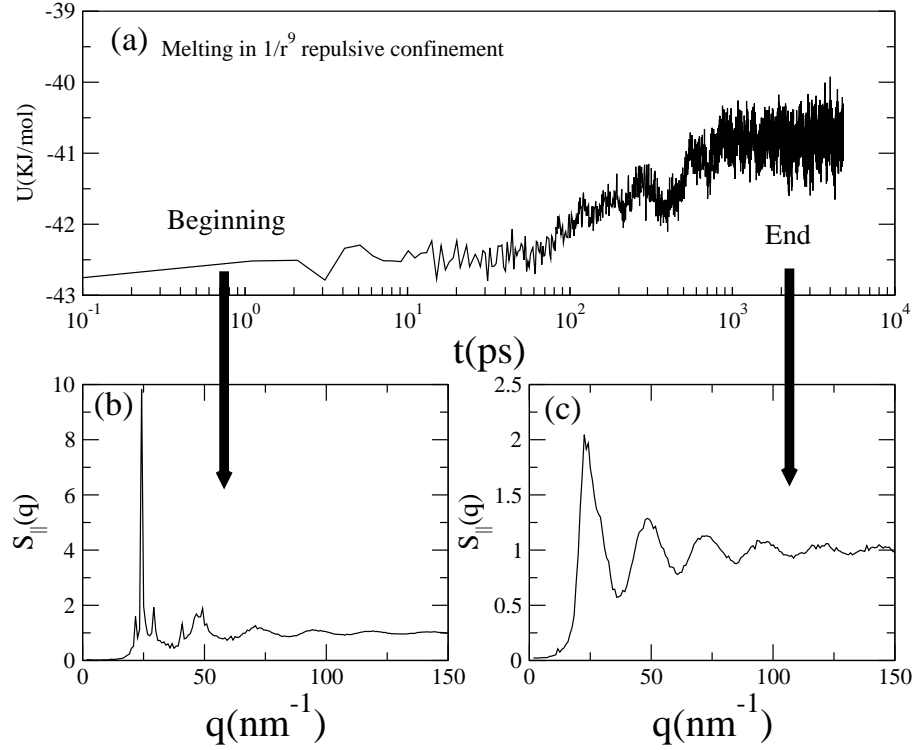}
\caption{(a) A plot of potential energy as a function of $t$, when
  the crystal formed in LJ confinement~\cite{kumar2005} is kept
  between the $1/r^9$ repulsive  wall. (b) The crystal structure
  indicated by the sharp Bragg peaks melts, and (c) turns into a
  liquid indicated by the absence of Bragg peaks.}
\label{fig:melting-repulsive}
\end{figure}

At plate separation of 1.1~nm with hydrophobic LJ confinement, water
crystallizes to trilayer ice~\cite{kumar2005}.  From our simulations
of TIP5P water in repulsive confinements with the same plate
separation of 1.1~nm, we find that the system does not freeze within
accessible simulation time scales for $1/r^9$ repulsive confinement;
however the system freezes for WCA confinement.  As a more stringent
confirmation of this fact, we also use a starting ice configuration
obtained from simulations with LJ confinement for the same thickness,
and confirm that the ice melts to a liquid with $1/r^9$ repulsive
confinement. In Fig.~\ref{fig:melting-repulsive}, we show the
evolution of potential energy and lateral structure factor with time,
when the crystal formed in LJ confinement~\cite{kumar2005} is kept
between the $1/r^9$ repulsive walls. The potential energy first
increases and then reaches its equilibrium value of the liquid
potential energy accompanied by a structural change from a crystal
(presence of sharp Bragg peaks) to a liquid (absence of Bragg peaks).

\begin{figure}[htb]
\includegraphics[width=12cm]{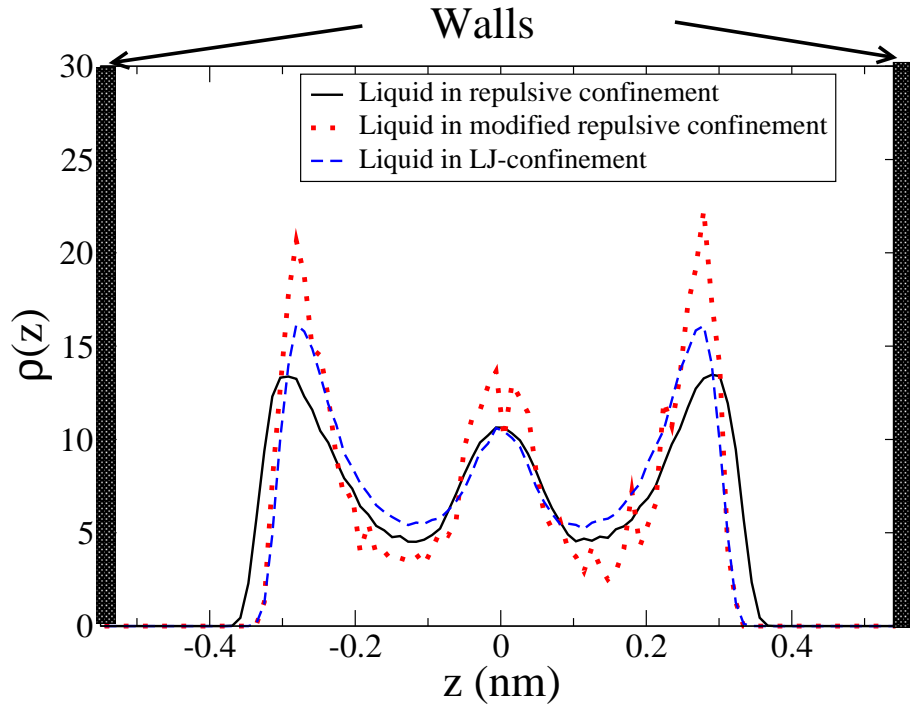}
\caption{Density profile $\rho (z)$ of water along z-direction for
different potentials at the same geometric density. The
repulsive confinement system freezes spontaneously when the parameters
of the potential are modified such that the effective $L_z$ calculated
from the $\rho (z)$ (red dotted line) for the repulsive system is same
as that for the LJ system (blue-dashed line) (see
Fig.~\ref{fig:potentials}).}
\label{fig:rhoz-comp}
\end{figure}

Based on this observation, it is tempting to claim that repulsion
inhibits crystallization, and that a preferable distance from the wall
determined by the attractive portion of the LJ potential is necessary
to induce crystallization.  However, as discussed above for the same
plate separation $1.1$~nm, $1/r^9$ repulsive confinement with the
chosen parameters increases the available space for molecules relative
to LJ confinement.  Hence to properly compare the crystallization
behavior, we must adjust the separation of the wall so that the
available space for the water molecules is the same in both systems.
We can make the available space the same by tuning the separation of
the plates or by tuning the potential. By tuning the parameters (see
Fig.~\ref{fig:potentials}, where the values of parameter
$\epsilon_{\rm OW}$ and $\sigma_{\rm OW}$ for the modified
$1/r^9$-repulsive potential are $1.25$~kJ/mol and $0.23$~nm
respectively) of the $1/r^9$repulsive potential in the repulsive
system such that the density profile along the $z$-axis becomes
similar, we have identical values of the available space between the
plates (see Fig.~\ref{fig:rhoz-comp}), and we find that an initial
crystal configuration does {\it not} melt, emphasizing that the
presence of the crystal is very sensitive to density {\it and} to
plate separation -- since the separation determines the accessible
packing arrangements between the plates.  Similar sensitivity to plate
separation for monolayer ice was seen in ref.~\cite{zangi2}.

In addition to examining the stability of initially crystalline
structures, we also consider whether freezing from the liquid state
occurs when we have the same effective plate separation.  We find that
for the repulsively confined systems, the crystal will also
spontaneously form if the available space between the plates is the
same as that for which initially crystalline configurations are stable.
Hence plate separation appears to be the dominant cause in
determining whether or not a crystal will form.


\section{Acknowledgments}

We thank F. Sciortino for helpful discussions, and the NSF for
support under grants CHE~0096892, CHE~0404699, and DMR-0427239. We
also thank the Boston University computational facility for
computational time and the Office of Academic Affairs at Yeshiva
University for support.

\newpage

\newpage

\newpage

\newpage

\newpage

\begin{figure}[htb]
\includegraphics[width=12cm]{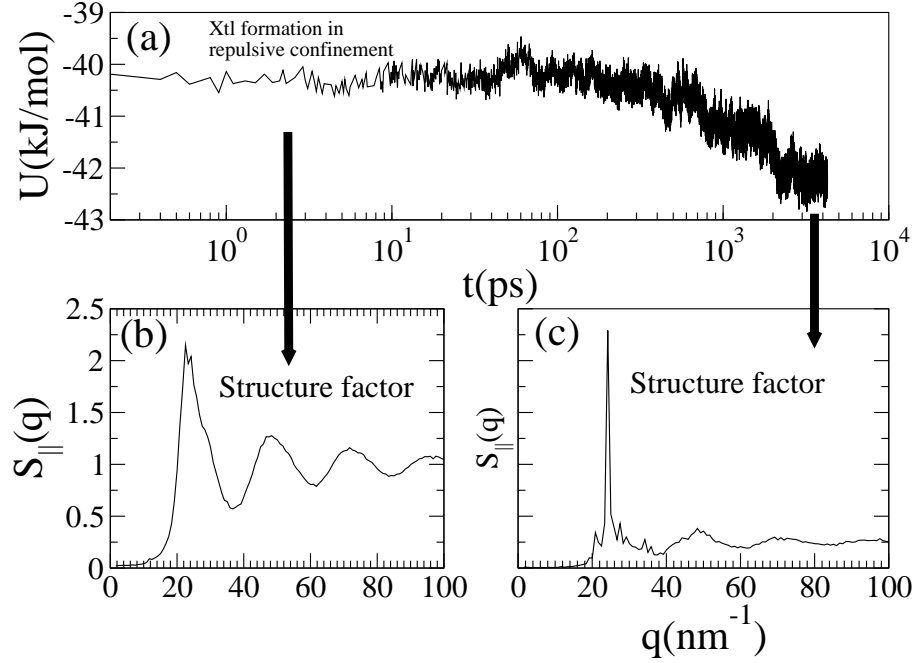}
\caption{(a) Potential energy as a function of time $t$, for $1/r^9$
  repulsive confinement when the effective $L_z$ is same as the
  effective $L_z$ of LJ confinement at $T=260$~K and geometric density
  $\rho_{\rm g} =0.981$~g/cm$^3$. The confined water spontaneously
  freezes, indicated by the drop in potential energy. (b) The
  structure factor of the ice such formed resembles the trilayer ice
  seen in case of LJ confinement ~\cite{zangi1,kumar2005}.}
\label{fig:xtal-repulsive}
\end{figure}

\newpage

\eject


\begin{thebibliography}{99}
\bibitem{note1} Two positive point charges of charge $q_H=0.241e$ (where $e$ is
the fundamental unit of charge) are located on each hydrogen atom at a
distance $0.09572$~nm from the oxygen atom; together they form an $HOH$
angle of $104.52^{\circ}$. Two negative point charges ($q_e=-q_H$)
representing the lone pair of electrons ($e^-$) are located at a
distance $0.07$~nm from the oxygen atom. These negative point charges
are located in a plane perpendicular to the $HOH$ plane and form an
$e^-Oe^-$ angle of $\cos^{-1}(1/3)=109.47^{\circ}$, the tetrahedral
angle.  To prevent overlap of molecules, a fifth interaction site is
located on the oxygen atom, and is represented by a LJ 
potential with parameters $\sigma_{OO}=0.312$~nm and
$\epsilon_{OO}=0.6694$~kJ/mol.
\end{thebibliography}
\end{document}